\documentclass[prd,onecolumn,12pt,tightenlines,superscriptaddress,nofootinbib,showpacs,shownokeys,floatfix]{revtex4} 

\usepackage{graphicx}
\usepackage{psfrag}
\usepackage{subfigure}

\begin{document}

\title{Search for very high energy gamma-rays from WIMP annihilations 
near the Sun with the Milagro Detector}

\author{R.~Atkins}
\altaffiliation{Department of Physics, University of Utah, 115 South 
1400 East, Salt Lake City, UT 84112}
\affiliation{Department of Physics, University of Wisconsin, 1150 
University Ave, Madison, WI 53706}

\author{W.~Benbow}
\altaffiliation{Max-Planck-Institut f\"ur Kernphysik, Postfach 103980, 
D-69029 Heidelberg, Germany}
\affiliation{Santa Cruz Institute for Particle Physics, University of 
California, 1156 High Street, Santa Cruz, CA 95064}

\author{D.~Berley}
\author{E.~Blaufuss}
\author{J.~Bussons}
\altaffiliation{Universit\'e de Montpellier II, Montpellier Cedex 5, 
F-34095, France}
\affiliation{Department of Physics, University of Maryland, College
Park, MD 20742}

\author{D.~G.~Coyne}
%R.~S.~Delay,\altaffilmark{\ref{uci}} 
\author{T.~DeYoung}
\affiliation{Department of Physics, University of Maryland, College
Park, MD 20742}

\author{B.~L.~Dingus}
\affiliation{Group P-23, Los Alamos National Laboratory, P.O. Box 1663, 
Los Alamos, NM 87545}

\author{D.~E.~Dorfan}
\affiliation{Santa Cruz Institute for Particle Physics, University of
California, 1156 High Street, Santa Cruz, CA 95064}

\author{R.~W.~Ellsworth}
\affiliation{Department of Physics and Astronomy, George Mason 
University, 4400 University Drive, Fairfax, VA 22030}

%A.~Falcone,\altaffilmark{\ref{unh}, 17} 

\author{L.~Fleysher}
\email{lazar.fleysher@physics.nyu.edu}
\author{R.~Fleysher}
\affiliation{Department of Physics, New York University, 4 Washington 
Place, New York, NY 10003}

\author{G.~Gisler}
\affiliation{Group P-23, Los Alamos National Laboratory, P.O. Box 1663,
Los Alamos, NM 87545}

\author{M.~M.~Gonzalez}
\affiliation{Department of Physics, University of Wisconsin, 1150
University Ave, Madison, WI 53706}

\author{J.~A.~Goodman}
\affiliation{Department of Physics, University of Maryland, College
Park, MD 20742}

\author{T.~J.~Haines}
\affiliation{Group P-23, Los Alamos National Laboratory, P.O. Box 1663,
Los Alamos, NM 87545}

\author{E.~Hays}
\affiliation{Department of Physics, University of Maryland, College
Park, MD 20742}

\author{C.~M.~Hoffman}
\affiliation{Group P-23, Los Alamos National Laboratory, P.O. Box 1663,
Los Alamos, NM 87545}

\author{L.~A.~Kelley}
\affiliation{Santa Cruz Institute for Particle Physics, University of
California, 1156 High Street, Santa Cruz, CA 95064}

\author{C.~P.~Lansdell}
\affiliation{Department of Physics, University of Maryland, College
Park, MD 20742}

\author{J.~T.~Linnemann}
\affiliation{Department of Physics and Astronomy, Michigan State 
University, 3245 BioMedical Physical Sciences Building, East Lansing, MI 
48824}

%J.~McCullough,\altaffilmark{\ref{ucsc},18} 

\author{J.~E.~McEnery}
\altaffiliation{NASA Goddard Space Flight Center, Greenbelt, MD 20771}
\affiliation{Department of Physics, University of Wisconsin, 1150
University Ave, Madison, WI 53706}

\author{R.~S.~Miller}
\affiliation{Department of Physics, University of New Hampshire, Morse 
Hall, Durham, NH 03824}

\author{A.~I.~Mincer}
\affiliation{Department of Physics, New York University, 4 Washington
Place, New York, NY 10003}

\author{M.~F.~Morales}
\altaffiliation{Massachusetts Institute of Technology, Building 37-664H, 
77 Massachusetts Avenue, Cambridge, MA 02139}
\affiliation{Santa Cruz Institute for Particle Physics, University of
California, 1156 High Street, Santa Cruz, CA 95064}

\author{P.~Nemethy}
\affiliation{Department of Physics, New York University, 4 Washington
Place, New York, NY 10003}

\author{D.~Noyes}
\affiliation{Department of Physics, University of Maryland, College
Park, MD 20742}

\author{J.~M.~Ryan}
\affiliation{Department of Physics, University of New Hampshire, Morse
Hall, Durham, NH 03824}

\author{F.~W.~Samuelson}
\affiliation{Group P-23, Los Alamos National Laboratory, P.O. Box 1663,
Los Alamos, NM 87545}

%M.~Schneider,\altaffilmark{\ref{ucsc}} 
%B.~Shen,\altaffilmark{16} 

\author{A.~Shoup}
\affiliation{Department of Physics and Astronomy, University 
of California, Irvine, CA 92697}

\author{G.~Sinnis}
\affiliation{Group P-23, Los Alamos National Laboratory, P.O. Box 1663,
Los Alamos, NM 87545}

\author{A.~J.~Smith}
\author{G.~W.~Sullivan}
\affiliation{Department of Physics, University of Maryland, College
Park, MD 20742}

%K.~Wang,\altaffilmark{14,19} 
%M.~Wascko,\altaffilmark{14,20} 

\author{D.~A.~Williams}
\author{S.~Westerhoff}
\altaffiliation{Department of Physics, Columbia University, 538 West 
120th Street,  New York, NY 10027}
\affiliation{Santa Cruz Institute for Particle Physics, University of
California, 1156 High Street, Santa Cruz, CA 95064}

\author{M.~E.~Wilson}
\affiliation{Department of Physics, University of Wisconsin, 1150
University Ave, Madison, WI 53706}

\author{X.~W.~Xu}
\affiliation{Group P-23, Los Alamos National Laboratory, P.O. Box 1663,
Los Alamos, NM 87545}

\author{G.~B.~Yodh}
\affiliation{Department of Physics and Astronomy, University
of California, Irvine, CA 92697}

\date{\today}

\begin{abstract}

The neutralino, the lightest stable supersymmetric particle, is a strong
theoretical candidate for the missing astronomical ``dark matter''.  A
profusion of such neutralinos can accumulate near the Sun when they lose
energy upon scattering and are gravitationally captured. 
Pair-annihilations of those neutralinos may produce very high energy
(VHE, above $100\ GeV$) gamma-rays.

Milagro is an air shower array which uses the water Cherenkov technique
to detect extensive air showers and is capable of observing VHE
gamma-rays from the direction of the Sun with an angular resolution of
$0.75^{\circ}$. Analysis of Milagro data with an exposure to the Sun of
1165 hours presents the first attempt to detect TeV gamma-rays produced
by annihilating neutralinos captured by the Solar system and shows no
statistically significant signal. Resulting limits that can be set on
gamma-ray flux due to near-Solar neutralino annihilations and on
neutralino cross-section are presented.

\end{abstract}


\maketitle

% ===================================================================

\section{Introduction}

There is very strong evidence that the Universe, and the galaxies in
particular, are full of non-baryonic ``dark matter'' (see, for example,
\cite{Rubin,netterfield,halverson,stompor,perlmutter,riess,spergel}). 
One candidate for this dark matter is the neutralino ($\chi$) --- a
weakly interacting massive particle (WIMP) predicted by super-symmetric
theories~\cite{Haber,Jungman}. Experimental tests for this possibility
include direct searches with extremely sensitive devices which can
detect energy deposited by a neutralino when it elastically scatters off
a nucleus and indirect searches which look for products of
neutralino-neutralino annihilations.

The Italian/Chinese collaboration (DAMA) employs an ionization bolometer
and has reported an observation which they interpret as consistent with
the annual modulation predicted if WIMPs exist~\cite{DAMA}. However,
there are possible modulating systematic errors. The CDMS experiment
detects phonon vibrations of the crystal lattice caused by the
neutralino-nucleon scattering in the detector volume and has obtained
data that appear to exclude most of the DAMA-allowed region~\cite{CDMS}.
They reach a spin-independent WIMP-nucleon cross-section limit around
$2\cdot 10^{-42}\ cm^{2}$ in the mass range $20-100\ GeV$. Edelweiss,
which utilizes both phonon and ionization bolometers, has also released
results that significantly cut into the DAMA allowed
region~\cite{Edelweiss}. 

% The summary of the limits of direct searches is shown in
% figure~\ref{fig:edelweiss}.

Indirect searches look for decay products from neutralino annihilations
coming from regions with enhanced neutralino densities. Searches vary by
the regions explored (such as the Galactic center or the Sun) and by the
decay products being detected. Unlike direct searches, interpretation of
these experiments requires assumptions about astrophysical parameters
which give rise to the neutralino annihilation distribution in the
region being studied.  They also depend on cross-sections and branching
ratios, both of which are supersymmetry-model dependent.

An increased density of neutralinos may exist in the vicinity of the
Galactic center or the Sun~\cite{PressSpergel}. This could have arisen
in the initial formation of these objects. Also, neutralinos entering
the Solar system (or the Galaxy) may lose energy via elastic scattering
with ordinary matter and become gravitationally trapped. Due to the
capture and repeated scatterings, there would be a near-solar (or
Galactic-center) enhancement in the neutralino density. Such a local
neutralino build-up may provide a detectable flux of annihilation
products. Examples of indirect searches are those by the Kamiokande and
SuperKamiokande underground neutrino detectors that have set limits on
solar and terrestrial neutralino-induced muon fluxes~\cite{Mori,
SuperK2001}. Such searches are helped by the ability of neutrinos to
escape from their production region and by the typically large predicted
yield, but are hindered by the small detection probability.

Another possible method for detecting dark matter particles is from
their annihilation into $\gamma$-rays. The Minimal Supersymmetric
extension of the Standard Model (MSSM) predicts that the gamma rays
emerging from $\chi\chi\rightarrow\gamma\gamma$ and $\chi\chi\rightarrow
Z\gamma$ neutralino annihilation modes will give distinct monochromatic
signals in the energy range between $100\ GeV$ and $10\ TeV$, depending
on the neutralino mass. An additional ``continuum'' spectrum signal of
photons will be produced by the decay of secondaries produced in the
non-photonic annihilation modes. However, past and present high energy
gamma-ray experiments, such as EGRET and the Whipple atmospheric
Cherenkov Telescope, lack the sensitivity to detect annihilation
$\gamma$-line fluxes predicted for many allowed supersymmetric models
and Milky Way halo profiles. The next generation ground-based and
satellite gamma-ray experiments, such as VERITAS and GLAST, will allow
exploration of portions of the MSSM parameter space, assuming that the
dark matter density is peaked at the galactic center~\cite{Bergstrom3}.

The Milagro $\gamma$-ray observatory, which has been taking data since
1999, is sensitive to cosmic gamma rays at energies around 1~TeV and is
capable of continuously monitoring the overhead sky with angular
resolution of $0.75^{\circ}$. In this paper, we present the results of a
search for a TeV gamma-ray signal from the vicinity of the Sun (1-2 
solar radii) with Milagro.

% ===================================================================

\section{Milagro detector}

The Milagro Extensive-Air Shower Array is located at $35.88^{\circ}$
North latitude and $106.68^{\circ}$ West longitude in the Jemez
Mountains near Los Alamos, New Mexico. At an altitude of $2630\ m$ above
the sea level, its atmospheric overburden is about $750\ g/cm^2$. The
Milagro detector, commissioned in June of 1999, records about 1700
extensive air shower events per second and is sensitive to gamma-showers
with energies above $\sim\!\! 100\ GeV$. The detector consists of a 21
kiloton water-filled pond instrumented with two layers of
photo-multiplier tubes (PMTs). These PMTs detect Cherenkov light
produced by secondary shower particles which enter the pond. The
top(bottom) layer has 450 (273) PMTs arranged on a $2.8\times 2.8\ m$
grid, $1.4\ (6)\ m$ below the water surface.

The direction of each shower is determined as the normal of the plane
fitted to the PMTs' times using an iterative weighted $\chi^{2}$-method
which rejects outliers. The weights for the $\chi^{2}$-fit are
prescribed based on the PMT signal strength. The angular resolution of
the array is estimated to be $0.75^{\circ}$.

Extensive air showers produced by cosmic rays are the primary source of
triggers in the experiment. These showers are likely to contain hadrons
that reach the ground level and produce hadronic cascades in the
detector, or muons that penetrate to the bottom layer, and will thus
illuminate a relatively small number of neighboring PMTs in that layer.
Photon induced showers, on the other hand, generally will produce rather
smooth light intensity distributions. Based on this observation, a
technique for identification of gamma versus hadron initiated showers
has been formulated~\cite{MilagroCrab} and according to computer
simulations can correctly select about 90\% of hadron initiated showers
and about 50\% of photon induced ones.

The fluctuations in the shower development, small size of the detector
and fluctuations in its response make energy determination on the
event-by-event basis difficult. The absolute energy scale can be
determined by examining the displacement of the shadow of the Moon due
to the Earth's magnetic field~\cite{MilagroMoon}. For a more detailed
description of the detector itself and reconstruction techniques used,
see references~\cite{milagritoNIM,milagroNIM,MilagroCrab}.

% ===================================================================

\section{Data analysis.}
\label{sec:analysis}

While it is difficult to tell the difference between cosmic-ray and
gamma-ray initiated showers, a gamma-ray signal can be detected as an
excess of events from the direction of the Sun above that expected from
the cosmic-ray background. The data analysis procedure thus entails
determining the average expected background signal $N_{on}^{b}$,
counting the number of events from the direction of the Sun $N_{on}$ and
determining the statistical significance of any excess found.

For each position of the Sun $N_{on}^{b}$ is found using event rates
from the same local region of the sky at a time when the Sun is not
present using the technique described in~\cite{RomanPaper}. This method
is based on isotropy of the cosmic-ray background and the assumption of
short time scale detector stability. It allows exclusion of known
sources from background estimation and correct restoration of the number
of excess events to be used for flux measurement.

The statistic $U$ chosen to test for an excess is:

\begin{equation}
  U=\frac{N_{on}-N_{on}^{b}}{\sqrt{\alpha N_{on}+N_{on}^{b}}}
\label{eq:LiMa}
\end{equation}

where $\alpha$ is the relative exposure ratio of the `` on-source''
region to the ``off-source'' one. For detailed discussion on how to deal
with varying $\alpha$ see~\cite{RomanPaper}. In the absence of a
source, $U$ approximately obeys a Gaussian distribution with zero mean
and unit variance.\footnote{The conditions of applicability of the
Gaussian approximation~\cite{RomanPaper} are satisfied for the
presented analysis~\cite{LazarPhD}.} If a source is present, $U$ will
still have an approximately Gaussian distribution with unit dispersion
but with a shifted mean.

For the current search we define at the outset the critical value of the
statistic $U$ as $u_{c}=5$ (which approximately corresponds to the level
of significance of $2.9\cdot 10^{-7}$). If no excess from the vicinity
of the Sun is found, we construct a limit on the source strength as a
strength which, if present, would have given us a detectable signal
($u>5$) with $97.7\%$ ($2\sigma$) probability~\cite{exclusionPaper}. In
addition, following the more standard procedure~\cite{eadie}, assuming
that the source exists, a $90\%$ one-sided confidence interval on its
strength is also calculated.

Given the detector response to particles of different types and assumed
source features, it is possible to predict the number of events $N_{k}$ 
to be observed due to the source. 

\begin{equation}
  N_{k}=\int F(k,E)S(k,\Theta)T(\Theta)
      \left[\int_{\tilde{\Theta}\in\tilde{\Omega}}
      A_{k}(E,\Theta,\tilde{\Theta})\,d\tilde{\Theta}\right] \,dE\,d\Theta
\label{eq:post_processing:N_Omega}
\end{equation}

where $F(k,E)$ is the number of particles of type $k$ with energy $E$
emitted by the source per unit area per unit time, $S(k,\Theta)$
describes the known geometrical shape of the source and is assumed to be
energy independent, $T(\Theta)$ is the time during which the source is
located in local direction $\Theta$ and $\tilde{\Theta}$ is the
direction of the particle arrival as output by the reconstruction
algorithm. The function $A_{k}(E,\Theta,\tilde{\Theta})$ is known as the
``effective area'' of the detector\footnote{The Milagro detector
response function $\left[\int_{\tilde{\Theta}\in\tilde{\Omega}}
A_{k}(E,\Theta,\tilde{\Theta})\,d\tilde{\Theta}\right]$ was obtained by
simulating the operation of Milagro detector with CORSIKA and GEANT
simulation packages (see~\cite{MilagroCrab}).} and can be integrated
over $\tilde{\Theta}$ prior to data processing for selected
configuration of $\tilde{\Omega}$. Therefore, after integration over
$\Theta$, the number of events due to particles of type $k$ to be
observed from a region $\tilde{\Omega}$ is:

\begin{equation}
        N_{k}=\int F(k,E)A_{k}^{T}(E)\,dE
\label{eq:Nevents_Flux}
\end{equation}

The integrated effective area $A_{k}^{T}(E)$ is obtained during the data
processing from Monte Carlo generated effective area
$A_{k}(E,\Theta,\tilde{\Theta})$.  By counting the number of excess
events in an observation bin $N_{\gamma}=N_{on}-N_{on}^{b}$, it is thus
possible to deduce some properties of the source function $F(\gamma,E)$.

% ===================================================================

\section{Effect of the Sun shadow}

The gamma ray signal from neutralino annihilations near the Sun should
appear as an excess number of events from the direction of the Sun over
the expected cosmic-ray background. The interpretation of any observed
signal, however, is not an easy problem. Largely, this is due to the
fact that the cosmic-ray background is not expected to be uniform; the
Sun absorbs the cosmic rays impinging on it and forms a cosmic-ray
shadow. The situation is complicated by the magnetic fields of the Earth
and the Sun. Due to bending of charged-particle trajectories in magnetic
fields, the Sun's shadow in the TeV range of particle energies will be
smeared and shifted from the geometrical position of the Sun. On the
other hand, in the presence of strong Solar magnetic fields, lower
energy particles cannot reach the surface of the Sun and are reflected
from it. Such particles are not removed from the interplanetary
medium and, since the cosmic rays are isotropically incident on the Sun,
may not even form a cosmic-ray shadow. In addition, the Solar magnetic
field varies with time, which will smear the shadow in a long-exposure
observation. Therefore, it is difficult to ascertain the exact shape of
the cosmic-ray shadow at the Sun's position and deduce an excess above
it.

Nevertheless, the effect of the Earth's magnetic field and the Solar
wind can be studied by observing the shadow of the Moon during the solar
day. The effect of the geomagnetic field on the shadows of the Sun and
the Moon should be very similar since the Sun and the Moon cover similar
size regions on the celestial sphere and traverse similar paths on the
local sky in one year of observation. In addition, the Earth's magnetic
field at the Moon distance is already so small that any additional
deflection beyond the Moon distance by this field of particles
originating from the Sun can be neglected.

A deficit of events due to such a shadow could be filled by a signal
from neutralinos which would remain undetected when the excess is
searched for. Since the shadow depth is unknown, when performing a
search for evidence of a positive excess we are forced to assume the
absence of the shadow. On the other hand, for setting a conservative
limit on the photon flux we use the maximum possible depth of the
shadow.

The deficit of events from the direction of the Sun can not be greater
than that produced by the Moon because as mentioned above the additional
Solar magnetic field can only decrease its shadow relative to the Moon.
To be conservative in setting the gamma-ray flux limit, the strongest
event deficit produced by the Moon in $5^{\circ}$ radius from its
position, corrected for relative Sun/Moon exposure, is used as a
correction for the possible presence of a shadow of the Sun.

% ===================================================================

\section{Results}
\subsection{Outcome of the observation}

\begin{figure}
\centering
%\begin{tabular}{cc}
%\includegraphics[width=2.7in]{FIGURES/moon_day_gp2.eps} &
%\includegraphics[width=2.7in]{FIGURES/sun_all_gp2.eps} \\
%\includegraphics[width=2.7in]{FIGURES/moon_day_exposure.eps} &
%\includegraphics[width=2.7in]{FIGURES/sun_exposure.eps} \\
%\end{tabular}
%\includegraphics[width=\textwidth]{FIGURES/significances.eps}
\subfigure[\label{subfig:moon_map}]{\includegraphics[width=2.7in]{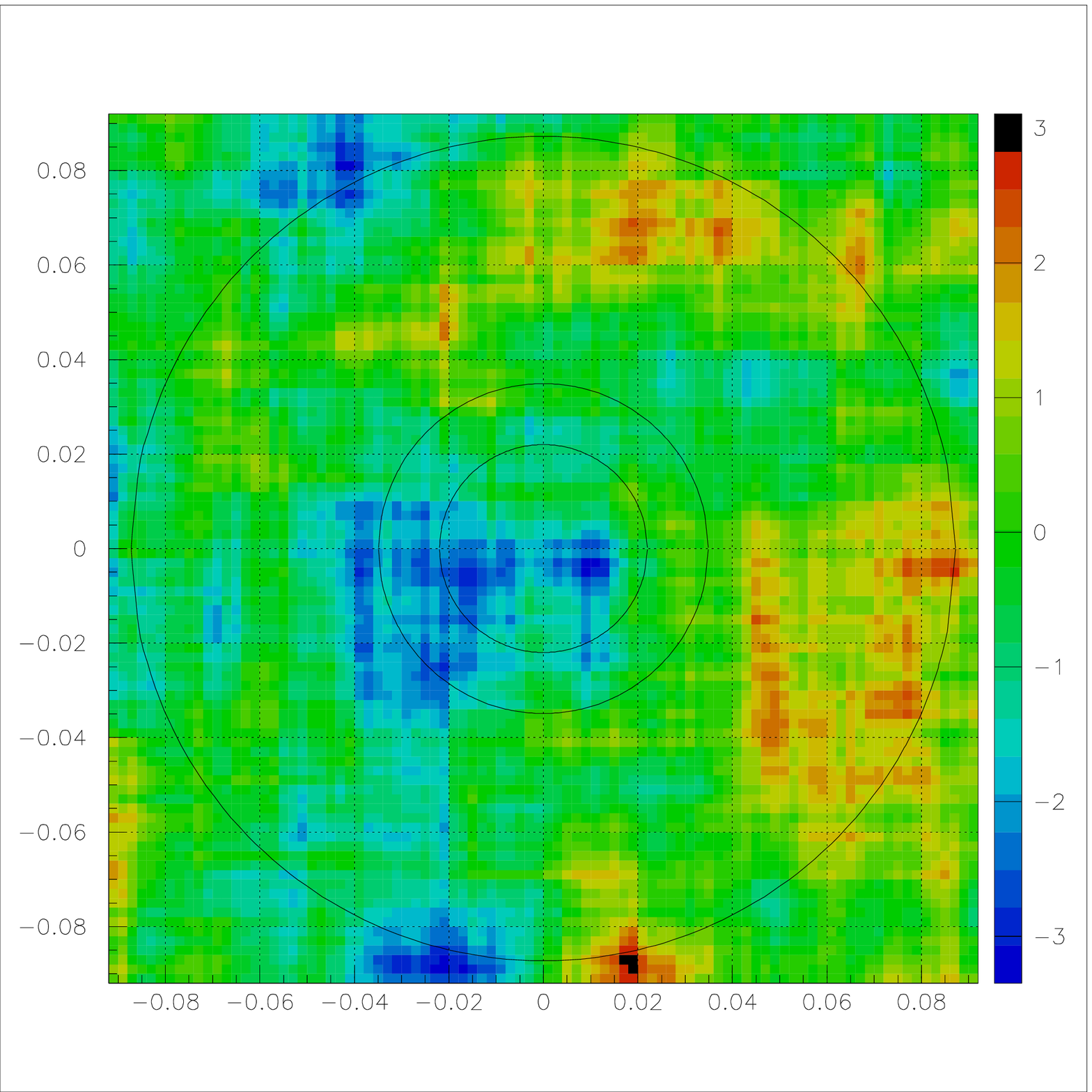}}
\subfigure[\label{subfig:sun_map}]{\includegraphics[width=2.7in]{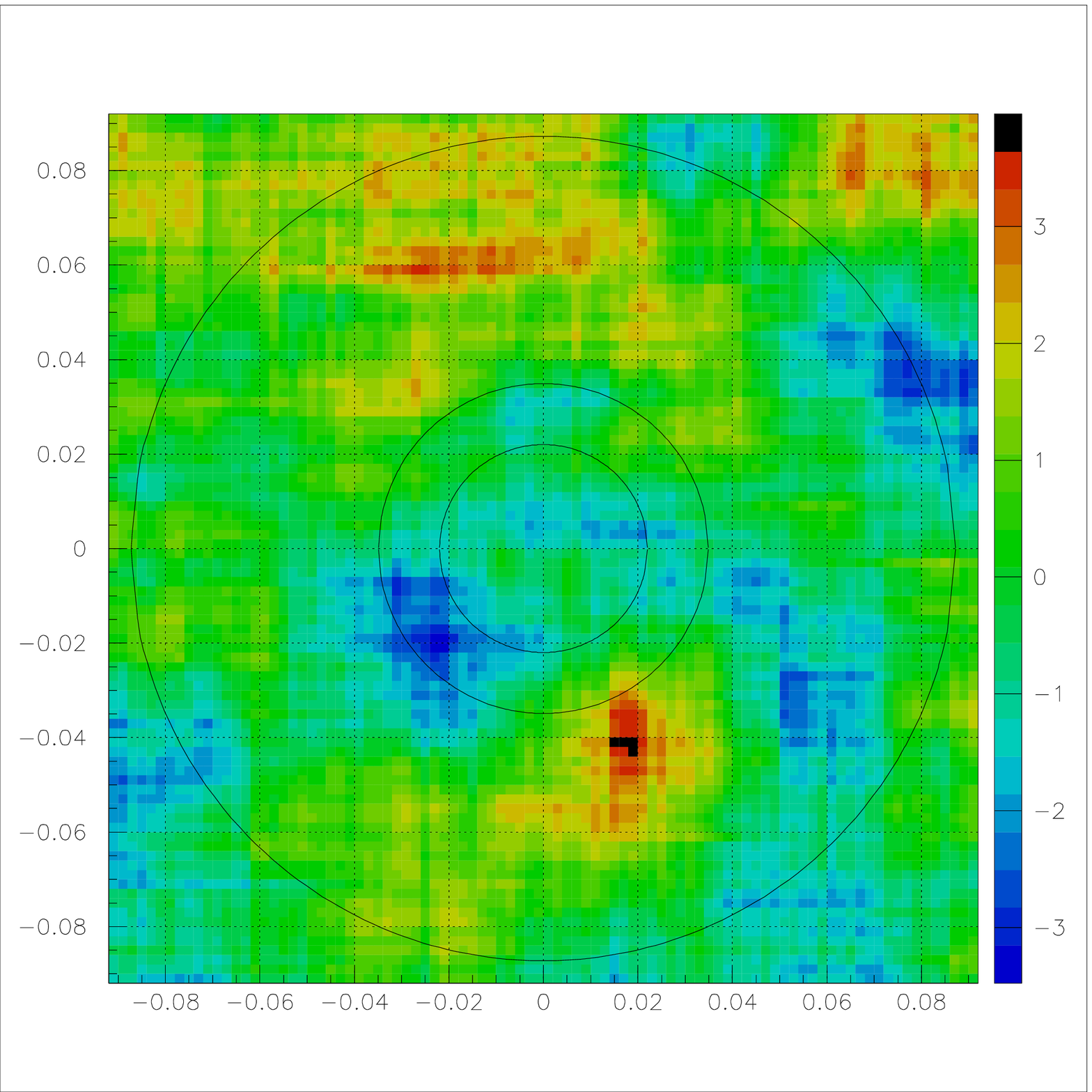}} \\
\subfigure[\label{subfig:moon_exp}]{\includegraphics[width=2.7in]{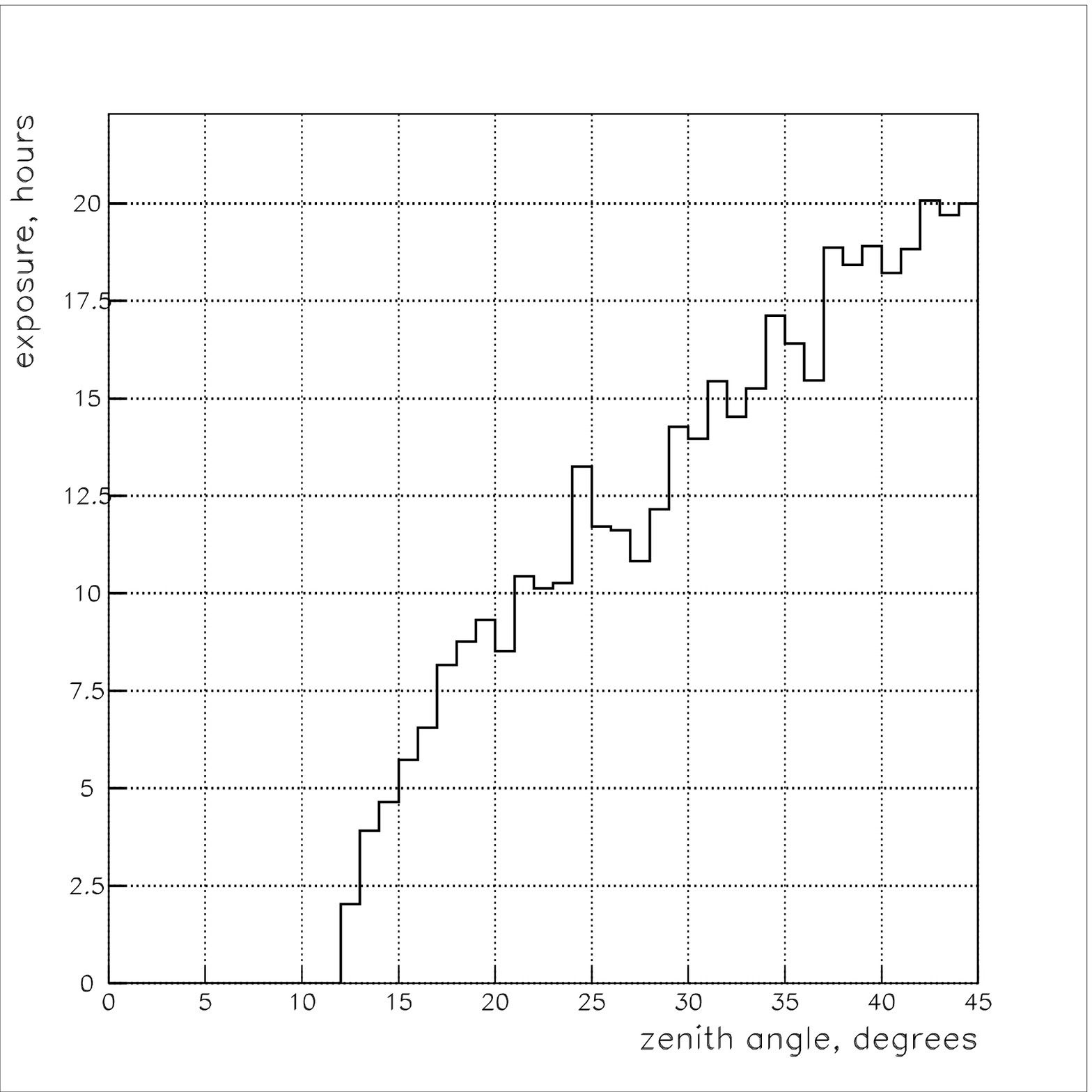}}
\subfigure[\label{subfig:sun_exp}]{\includegraphics[width=2.7in]{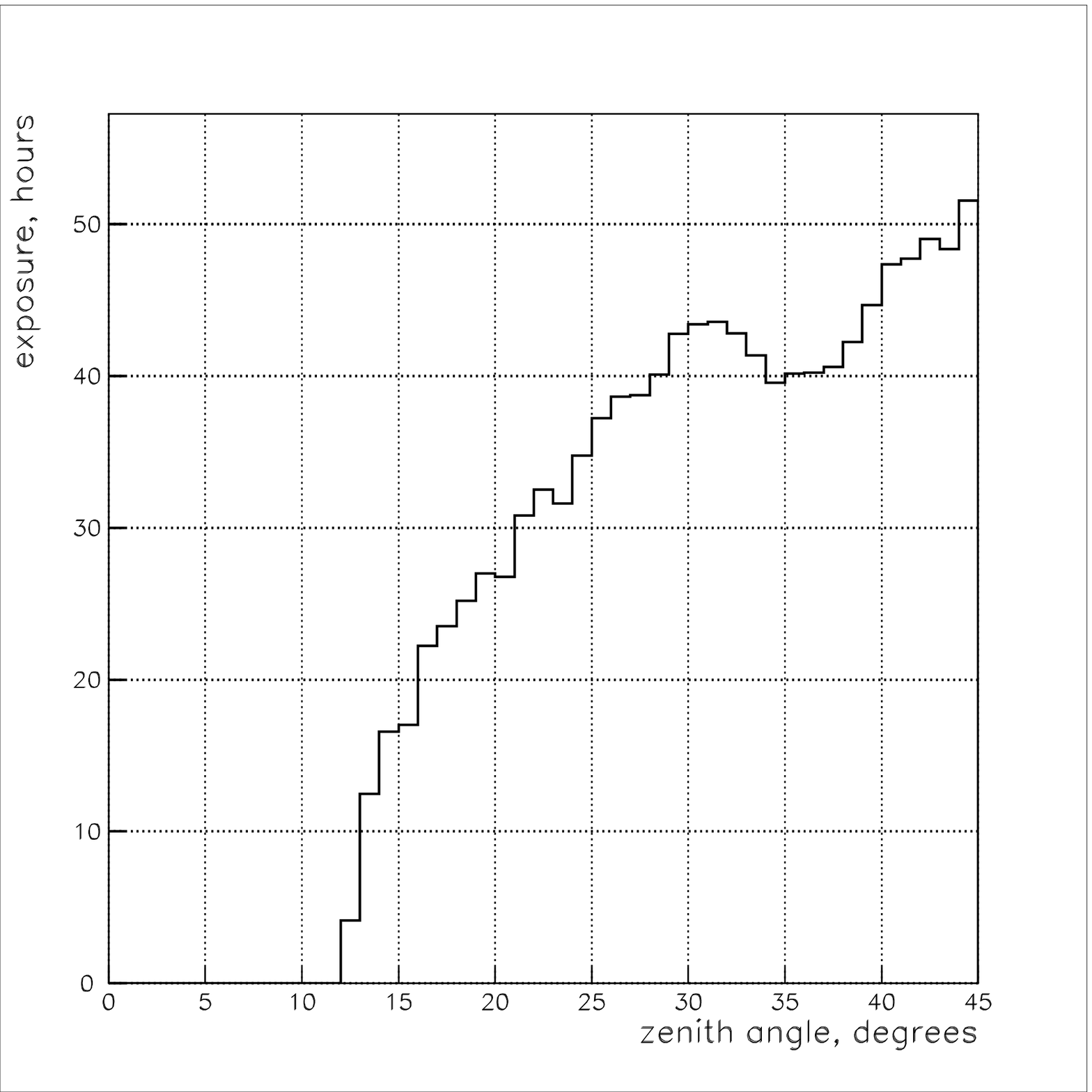}}
\caption{The maps of statistical significance of the number of excess 
         events from the regions of the sky around the daytime 
         Moon (\ref{subfig:moon_map}) and the Sun (\ref{subfig:sun_map}) 
         and the corresponding source exposure as function of zenith 
         angle in hours per degree (\ref{subfig:moon_exp},\ref{subfig:sun_exp}). 
         The color code is the value of $U$ 
         (see equation~(\ref{eq:LiMa})). The Moon shadow is much
         less significant than that reported 
         in~\cite{MilagroMoon} because of $\gamma$-selection and 
         other cuts used in the present 
         analysis. The maps are made using azimuthal equal-area 
         projection in polar case centered on the corresponding 
         Celestial object. Contours represent loci of distance 
         $1.26^{\circ}$, $2.0^{\circ}$ and $5.0^{\circ}$ from the 
         Celestial object.
\label{fig:maps_and_exposure}}
\end{figure}

The data used in this work was chosen to satisfy the following
conditions: online reconstruction between the 19th of July 2000 and the
10th of September 2001, the number of PMTs required for a shower to
trigger the detector greater than $60$, the number of PMTs used in the
angular reconstruction greater than $20$, zenith angles smaller than
$45$ degrees, and passing the gamma/hadron separation
cut~\cite{MilagroCrab}. The start and end dates correspond respectively
to introduction of the hadron separation parameter into the online
reconstruction code and detector turn-off for scheduled repairs. Several
data runs were removed from the dataset which included calibration runs
and the data recorded when there were online DAQ problems.

For the solar region analysis, $\pm 5^{\circ}$ regions around the Moon
and the Crab nebula were vetoed from the data set as they present known
sources of anisotropy to the cosmic-ray background.

Photons produced in the Sun will be absorbed, whereas the distribution
of neutralino annihilations outside is a rapidly falling function of
distance from the Sun. Therefore, we believe that the gamma-ray signal
is produced mostly between 1 and 2 solar radii and treat the gamma-ray
source as circle of $0.5^{\circ}$ radius. It has been
shown~\cite{LazarPhD} that the optimal bin size is a slow function of
the source size and for estimated $0.75^{\circ}$ angular resolution of
the detector, the optimal ``on-source'' bin is a circular one with the
radius of $1.26^{\circ}$ centered on the Sun.

Overall, $1164.7$ hours of exposure to the Sun is obtained in the data
set. The total number of events observed in the ``on-source'' bin is
$N_{on}=137211$ while $N_{on}^{b}=137728$ events is expected based on
the ``off-source'' exposure, leading to the value of the test statistic
$U$ of $-1.35$ (see figure~\ref{fig:maps_and_exposure}). Therefore, the
null hypothesis of the absence of gamma-ray emission from the Sun can
not be rejected and a limit on the possible $\gamma$-ray flux from the
solar region is obtained.

Overall, 423.5 hours of exposure to the Moon during the day time is
obtained in this data set after $\pm 5^{\circ}$ regions around the Sun
and the Crab nebula were vetoed. The largest deficit observed in the sky
map centered on the geometrical position of the Moon is $-3.3$
corresponding to $-766$ events (see
figure~\ref{fig:maps_and_exposure}). The exposure on the Sun is $2.75$
times greater than that on the Moon during solar day, leading to
estimated maximal deficit in the Sun's direction as:
$-766\cdot 2.75=-2107$.

Hence, using the criteria outlined in section~\ref{sec:analysis} we
conclude that the number of excess events due to a possible Solar source
of VHE photons does not exceed $N_{exclude}=4791$. It is this number
which is used to construct the exclusion region on gamma-ray flux
strength from the Solar region (equation~\ref{eq:Nevents_Flux}).

The $90\%$ one-sided confidence interval on the average number of excess
events $N_{confidence}$ assuming that gamma-ray emission from the Sun
exists constructed in a standard way~\cite{eadie} is
$N_{confidence}=2081$. It is this number which would allow estimation of
the gamma-ray flux strength from the Solar region if it were known that
such gamma-ray emission exists (equation~\ref{eq:Nevents_Flux}).

% ===================================================================

\subsection{Limit on the photon flux due to neutralino annihilations}

\begin{figure}
\centering
\includegraphics[width=0.7\columnwidth]{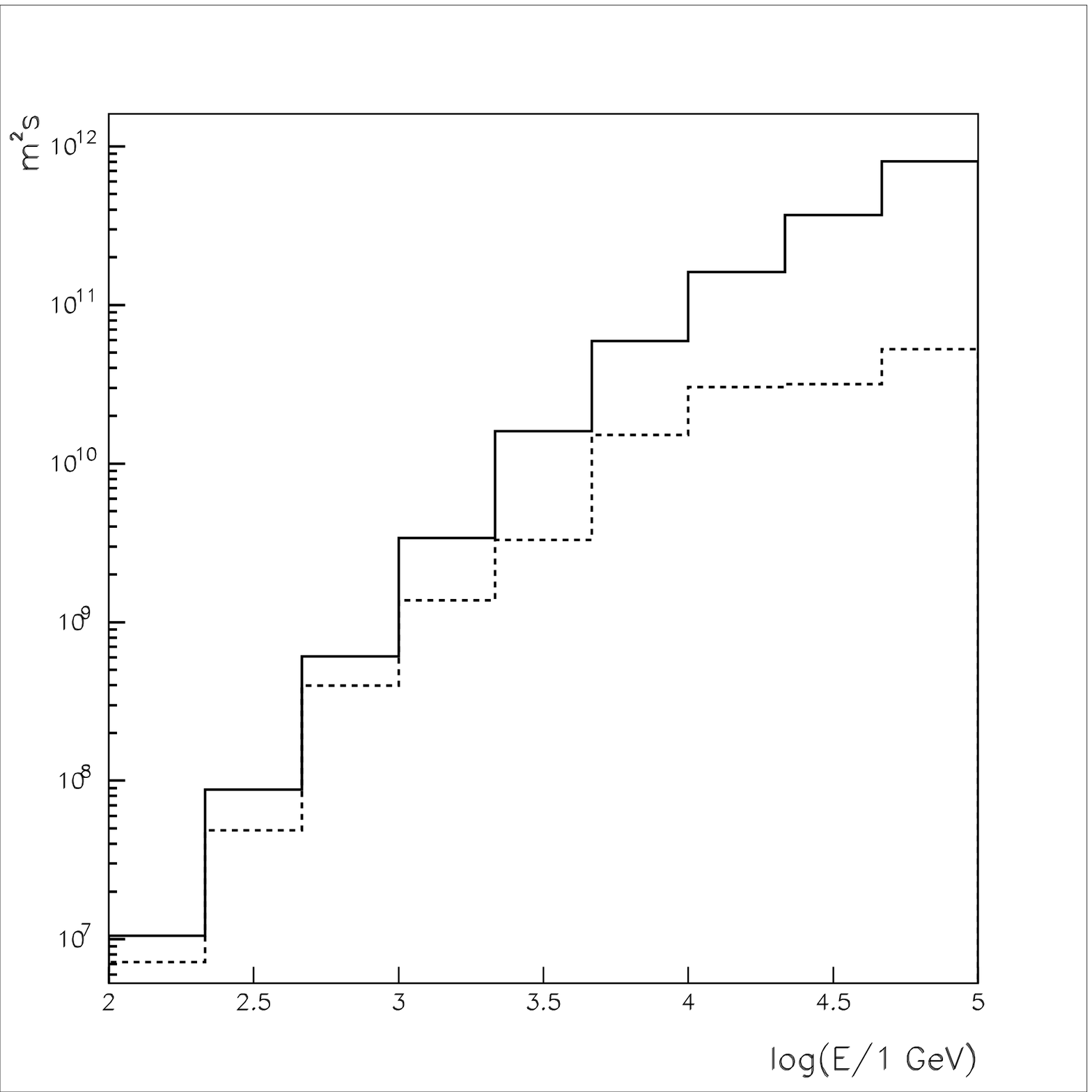}
\caption{Integrated effective area $A_{k}^{T}$ as a function of photon
         (solid) and proton (dashed) energy (see 
         equation~(\ref{eq:Nevents_Flux})) for the circular 
         $1.26^{\circ}$ bin centered on the Sun for the trigger 
         condition and cuts used in this analysis.
         \label{fig:effective_area}}
\end{figure}

\begin{figure}
\begin{center}
%\begin{tabular}{cc}
%\includegraphics[width=0.45\textwidth]{FIGURES/flux_limits_02_1.eps} &
%\includegraphics[width=0.45\textwidth]{FIGURES/flux_limits_2_50.eps} 
%\end{tabular}
%\plottwo{FIGURES/flux_limits_02_1.eps}{FIGURES/flux_limits_2_50.eps}
\subfigure{\includegraphics[width=0.45\columnwidth]{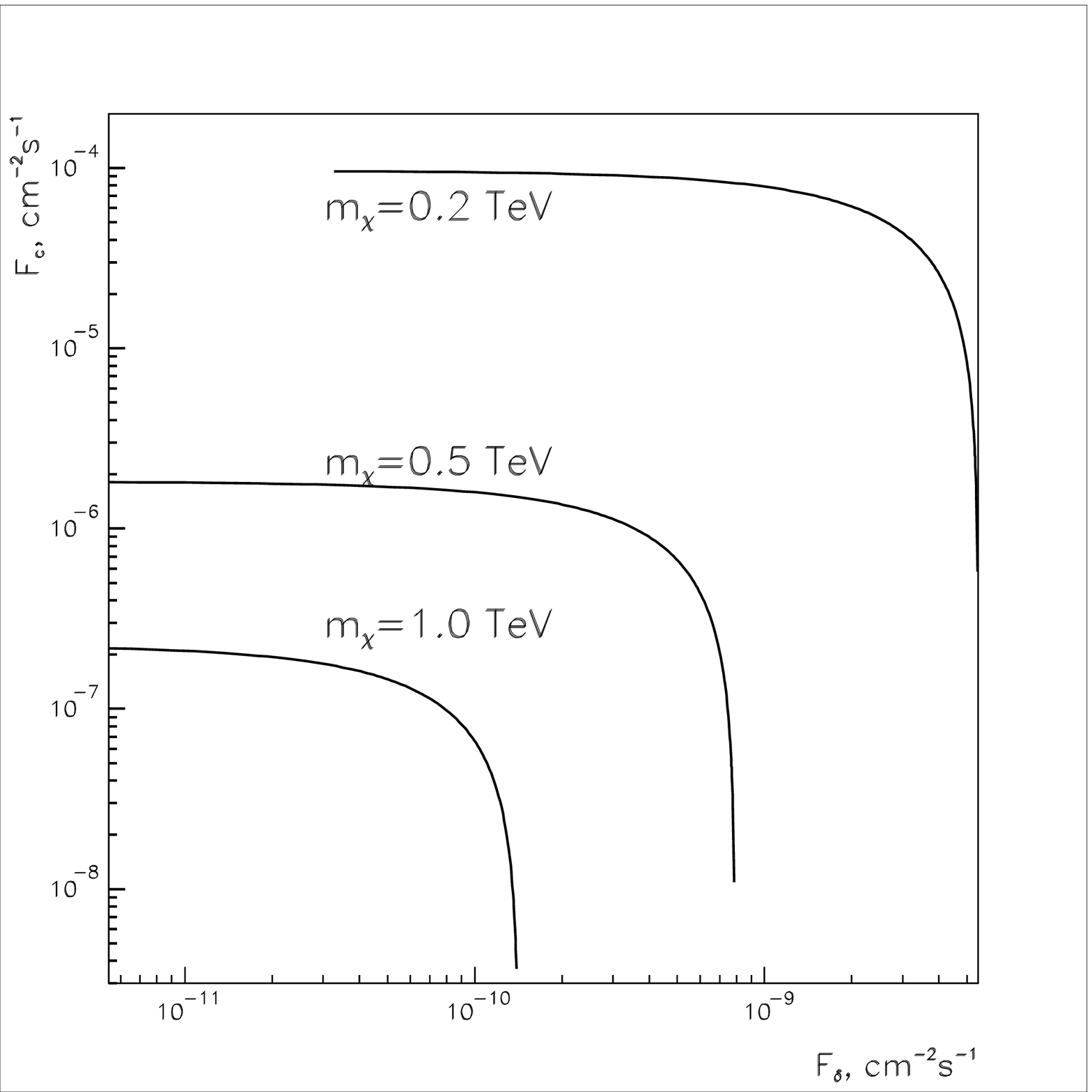}}
\subfigure{\includegraphics[width=0.45\columnwidth]{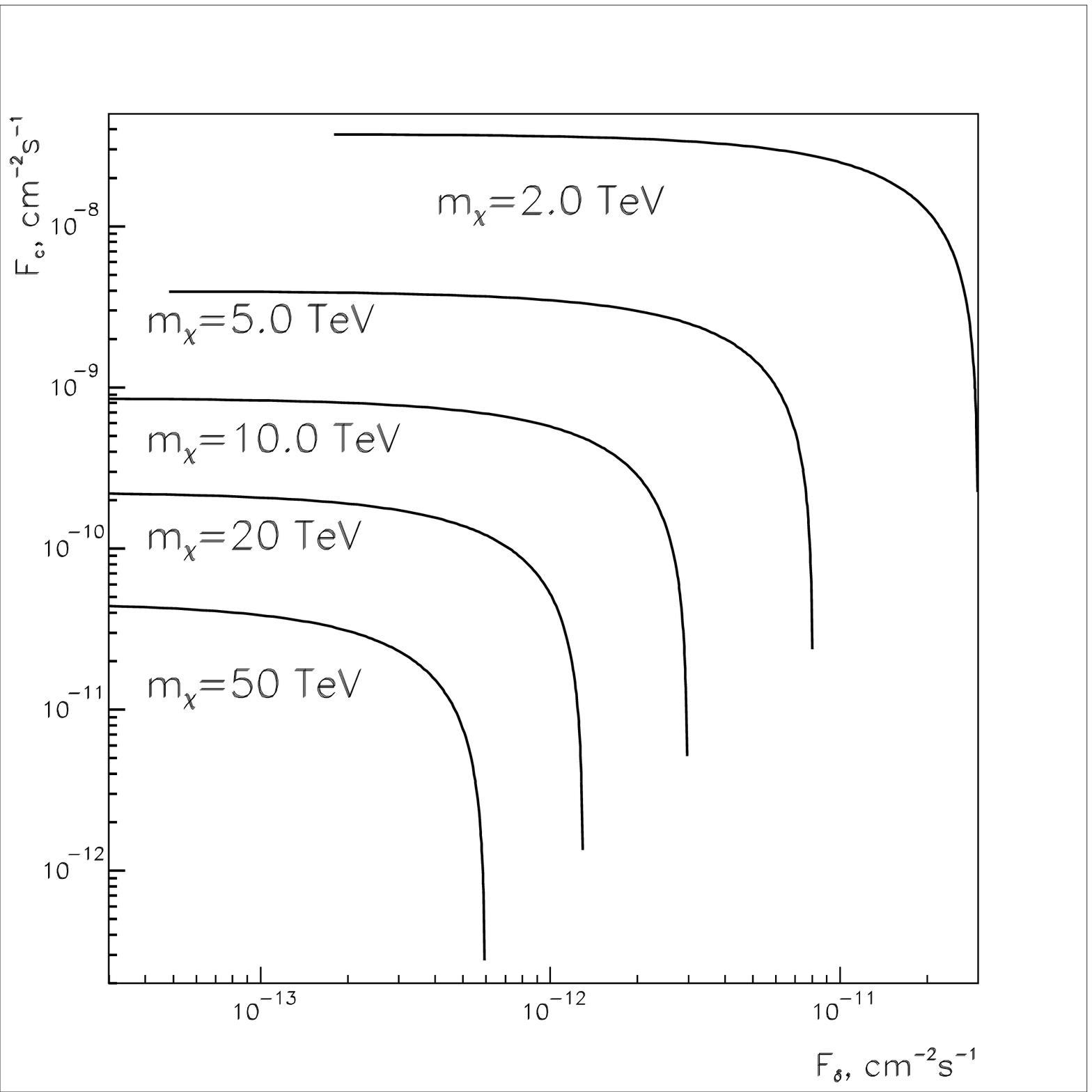}}
\end{center}
\vspace{-0.5pc}
\caption{The values of ($F_{\delta}$,$F_{c}$) below the lines are 
         allowed for corresponding neutralino masses.
         \label{fig:flux_constraints}}
\end{figure}

\begin{table}
\centering
\begin{tabular}{||c||c||c||} \hline\hline
$m_{\chi}\ (TeV)$ & $\Delta\ (cm^{2}s)$ & $\Sigma\ (cm^{2}s)$ \\ 
\hline\hline
0.1  & $1.055\times 10^{11}$  &  $0.000$             \\ \hline
0.2  & $8.772\times 10^{11}$  &  $4.969\times 10^{7}$ \\ \hline
0.5  & $6.070\times 10^{12}$  &  $2.634\times 10^{9}$ \\ \hline
1.0  & $3.389\times 10^{13}$  &  $2.127\times 10^{10}$ \\ \hline
2.0  & $1.600\times 10^{14}$  &  $1.280\times 10^{11}$ \\ \hline
5.0  & $5.942\times 10^{14}$  &  $1.208\times 10^{12}$ \\ \hline
10.0 & $1.608\times 10^{15}$  &  $5.575\times 10^{12}$ \\ \hline
20.0 & $3.684\times 10^{15}$  &  $2.136\times 10^{13}$ \\ \hline
50.0 & $8.030\times 10^{15}$  &  $1.035\times 10^{14}$ \\ \hline \hline
\end{tabular}
\caption{Coefficients of the flux limit calculation (see 
         equation~(\ref{eq:flux_constraints})).
         \label{table:flux_limit_coef}}
\end{table}

\begin{table}
\centering
\begin{tabular}{||c||c||c||}              \hline \hline
$m_{\chi}\ (TeV)$ & $F_{\delta}<\ (cm^{-2}s^{-1})$ & 
                          $F_{c}<\ (cm^{-2}s^{-1})$  \\ \hline 
\hline
0.1    & $4.54\times 10^{-8}$  &  ---    \\ \hline
0.2    & $5.46\times 10^{-9}$  & $9.64\times 10^{-5}$    \\ \hline
0.5    & $7.89\times 10^{-10}$ & $1.82\times 10^{-6}$    \\ \hline
1.0    & $1.41\times 10^{-10}$ & $2.25\times 10^{-7}$    \\ \hline
2.0    & $2.99\times 10^{-11}$ & $3.74\times 10^{-8}$    \\ \hline
5.0    & $8.06\times 10^{-12}$ & $3.97\times 10^{-9}$    \\ \hline
10.0   & $2.98\times 10^{-12}$ & $8.59\times 10^{-10}$    \\ \hline
20.0   & $1.30\times 10^{-12}$ & $2.24\times 10^{-10}$  \\ \hline
50.0   & $5.97\times 10^{-13}$ & $4.63\times 10^{-11}$ \\ \hline\hline
\end{tabular}
\caption{The limit on the flux parameters $(F_{\delta},F_{c})$
         corresponding to the significance $2.9\cdot 10^{-7}$ and the 
         power of the test of $97.7\%$. For the $90\%$ 
         one-sided confidence interval multiply these limits by 
         $0.4344$.
         \label{table:flux_constraints}}
\end{table}

Because the two close direct-production spectral lines can not be
resolved by the Milagro detector, the differential photon flux due to
neutralino annihilations is assumed to have the form
(see~\cite{Bergstrom3}):

\begin{equation}
   \frac{dF(E_{\gamma})}{dE_{\gamma}}=F_{\delta}\delta(E_{\gamma}-m_{\chi})+
   \frac{F_{c}}{m_{\chi}}\cdot
      \frac{\Big(\frac{E_{\gamma}}{m_{\chi}}\Big)^{-3/2}e^{-7.8E/m_{\chi}}}%
                                {\int_{0.01}^{1}x^{-3/2}e^{-7.8x}dx}
\label{eq:Begstrom_flux}
\end{equation}

where $F_{\delta}$ is the integral flux due to a $\delta$-function-like
photon annihilation channel and $F_{c}$ is the integral flux of photons
with energies greater than $0.01\cdot m_{\chi}$ due to continuum photon
spectrum annihilation channel of neutralinos with mass $m_{\chi}$. Here,
the normalization of the continuum photon spectrum has been written out
explicitly.

Using this expression for the flux with the integrated effective area of
the detector, we find a relationship between the number of observed
events $N_{\gamma}$ and the integral flux parameters $F_{\delta}$ and
$F_{c}$ in the form:

\begin{equation}
  F_{\delta}\Delta + F_{c}\Sigma=N_{\gamma}
\label{eq:flux_constraints}
\end{equation}

The values of coefficients $\Delta$ and $\Sigma$ (see table
\ref{table:flux_limit_coef}) are obtained by substituting the expression
for flux~(\ref{eq:Begstrom_flux}) and the integrated effective area
(figure~\ref{fig:effective_area}) into the formula for the number of
expected events~(\ref{eq:Nevents_Flux}).\footnote{Should an emission
model other than the one presented by equation~(\ref{eq:Begstrom_flux})
become a theoretical preference, figure~\ref{fig:effective_area} can be
used to recompute the model's parameters.} Note, that depending on 
$m_{\chi}$ the integral in~(\ref{eq:Nevents_Flux}) may extend below the 
$0.01m_{\chi}$ normalization.

The figure~\ref{fig:flux_constraints} presents the curves demarcating
the allowed and excluded regions in the photon flux parameter space
$(F_{\delta},F_{c})$ corresponding to the significance $2.9\cdot
10^{-7}$ and the power of the test of $97.7\%$ obtained by setting
$N_{\gamma}=N_{exclude}$ which have a form of straight lines in the
$(F_{\delta},F_{c})$ plane. Because there is only one equation
(eq.~(\ref{eq:flux_constraints})) constraining two parameters,
table~\ref{table:flux_constraints} provides the most conservative limits
on each of $F_{\delta}$ and $F_{c}$ when the contribution from the other
is set to zero.

For the depiction of the one-sided $90\%$ confidence interval in the
$(F_{\delta},F_{c})$ plane, both axes in the
figure~\ref{fig:flux_constraints} should be rescaled by
$N_{exclude}/N_{confidence}=0.4344$.
%%%%%%%%%$N_{exclude}/N_{confidence}=2081/4791$.

Figure~\ref{fig:flux_constraints} is the derived limit on the values of
the parameters of the gamma-ray emission model
(equation~(\ref{eq:Begstrom_flux})) due to near solar WIMP annihilation
and is independent of the models of their distribution in the Milky Way
galaxy and the Solar system.

% ===================================================================

\subsection{Neutralino limits\label{sec:results_neutralino}}

\begin{table}
\centering
\begin{tabular}{||c||c||}           \hline \hline
$m_{\chi},\ (TeV)$  &
    $I(m_{\chi})\times\frac{10^{-43}cm^{2}}{\sigma_{p\chi}}
                   \frac{0.3GeV/cm^{3}}{\rho_{0}},\ (s^{-1})$
                                                        \\  \hline\hline
 $0.1$  & $1.65\times 10^{18}$   \\ \hline
 $0.2$  & $4.17\times 10^{17}$   \\ \hline
 $0.5$  & $6.72\times 10^{16}$   \\ \hline
 $1.0$  & $1.68\times 10^{16}$  \\ \hline
 $2.0$  & $4.22\times 10^{15}$  \\ \hline
 $5.0$  & $6.72\times 10^{14}$  \\ \hline
$10.0$  & $1.69\times 10^{14}$  \\ \hline
$20.0$  & $4.22\times 10^{13}$  \\ \hline
$50.0$  & $6.75\times 10^{12}$  \\ \hline\hline
\end{tabular}
\caption{Capture rate $I(m_{\chi})$ as a function of neutralino mass 
         $m_{\chi}$ obtained in a three-dimensional 
         calculation~\cite{LazarPhD}. The normalizations of 
         $\sigma_{p\chi}$ and $\rho_{0}$ are typically used numbers.
         \label{table:predicted_capture}}
\end{table}

\begin{table}
\centering
\begin{tabular}{||c||c||c||}              \hline \hline
$m_{\chi}\ (TeV)$ & 
        $\left(J/I\right)\frac{\sigma_{p\chi}}{10^{-43}cm^{2}}
        \frac{\rho_{0}}{0.3GeV/cm^{3}}b_{\gamma}^{c}f_{out}<$ 
& 
        $\left(J/I\right)\frac{\sigma_{p\chi}}{10^{-43}cm^{2}}
         \frac{\rho_{0}}{0.3GeV/cm^{3}}b_{\gamma}^{\delta}f_{out}<$ 
\\ \hline \hline
0.1    &  ---               & $155$      \\ \hline
0.2    & $1.30\times 10^{6}$ & $73.7$    \\ \hline
0.5    & $1.52\times 10^{5}$ & $66.1$    \\ \hline
1.0    & $7.54\times 10^{4}$ & $47.3$    \\ \hline
2.0    & $4.99\times 10^{4}$ & $39.9$    \\ \hline
5.0    & $3.32\times 10^{4}$ & $67.5$    \\ \hline
10.0   & $2.86\times 10^{4}$ & $99.2$    \\ \hline
20.0   & $2.99\times 10^{4}$ & $173$     \\ \hline
50.0   & $3.86\times 10^{4}$ & $497$     \\ \hline\hline
\end{tabular}
\caption{The limits on 
         $J(m_{\chi})/I(m_{\chi})\sigma_{p\chi}\rho_{0}b_{\gamma}^{c}f_{out}$
         and
         $J(m_{\chi})/I(m_{\chi})\sigma_{p\chi}\rho_{0}b_{\gamma}^{\delta}f_{out}$.
         \label{table:neutralino_limits}}
\end{table}

\begin{figure}
\centering
\includegraphics[width=0.7\columnwidth]{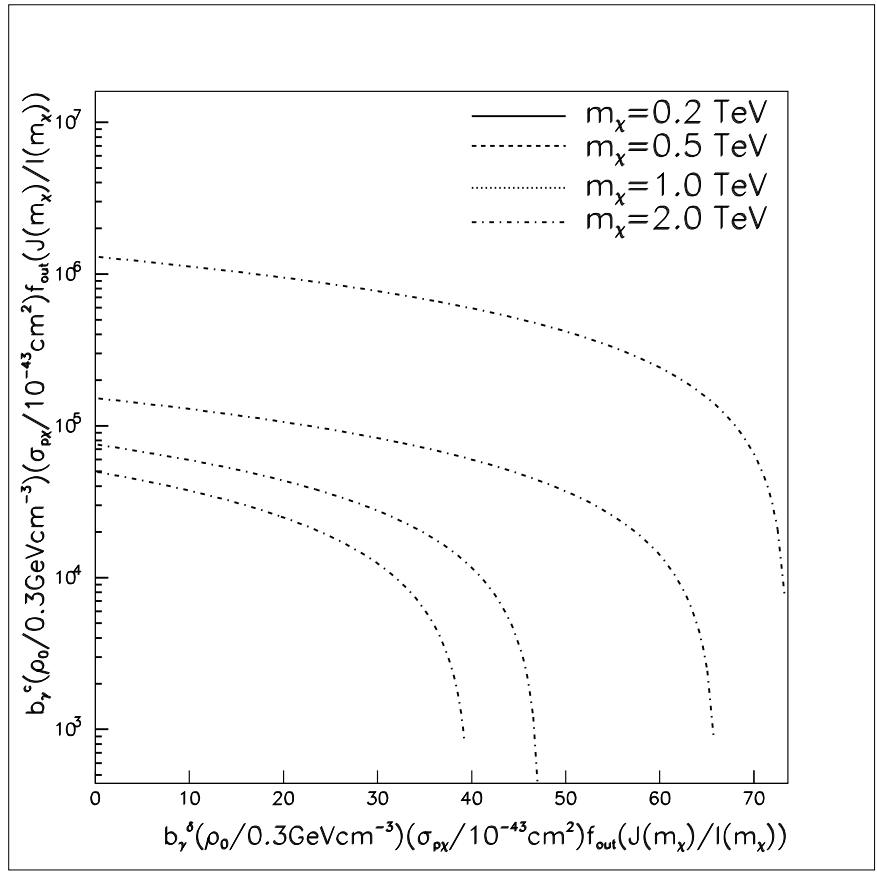}
%\plotone{FIGURES/cross4.eps}
\caption{The values of 
        $\Big(\rho_{0}\sigma_{p\chi}b_{\gamma}^{\delta}f_{out}
        J(m_{\chi})/I(m_{\chi})$,
        $\rho_{0}\sigma_{p\chi}b_{\gamma}^{c}f_{out}J(m_{\chi})/I(m_{\chi})\Big)$ 
        below the 
        lines are allowed based the constructed limit for corresponding 
        neutralino masses. As one progresses from low to high $m_{\chi}$, 
        the detector effective area goes up at the same time the flux of 
        incoming neutralinos goes down as $1/m_{\chi}$ for fixed dark 
        matter density $\rho_{0}$. In addition, the capture probability 
        is also decreasing as $1/m_{\chi}$ in an elastic scattering on a 
        fixed mass target. This explains qualitative behavior of these 
        lines.
        \label{fig:neutralino_limits}}
\end{figure}

\begin{figure}
\centering
%\psfrag{fbJ}{\Big[$\mathbf{\sigma_{p\chi}f_{out}b_{\gamma}^{\delta}J(m_{\chi})/I(m_{\chi})}$\Big]}
%\includegraphics[width=0.7\columnwidth]{FIGURES/my_edelweiss.eps}

\psfrag{bd}{$b_{\gamma}^{\delta}=10^{-3},\ b_{\gamma}^{c}=0$}
\psfrag{bc}{$b_{\gamma}^{\delta}=0,\ b_{\gamma}^{c}=1$}
\psfrag{fout}{$\times f_{out}$}

\includegraphics[width=0.7\columnwidth]{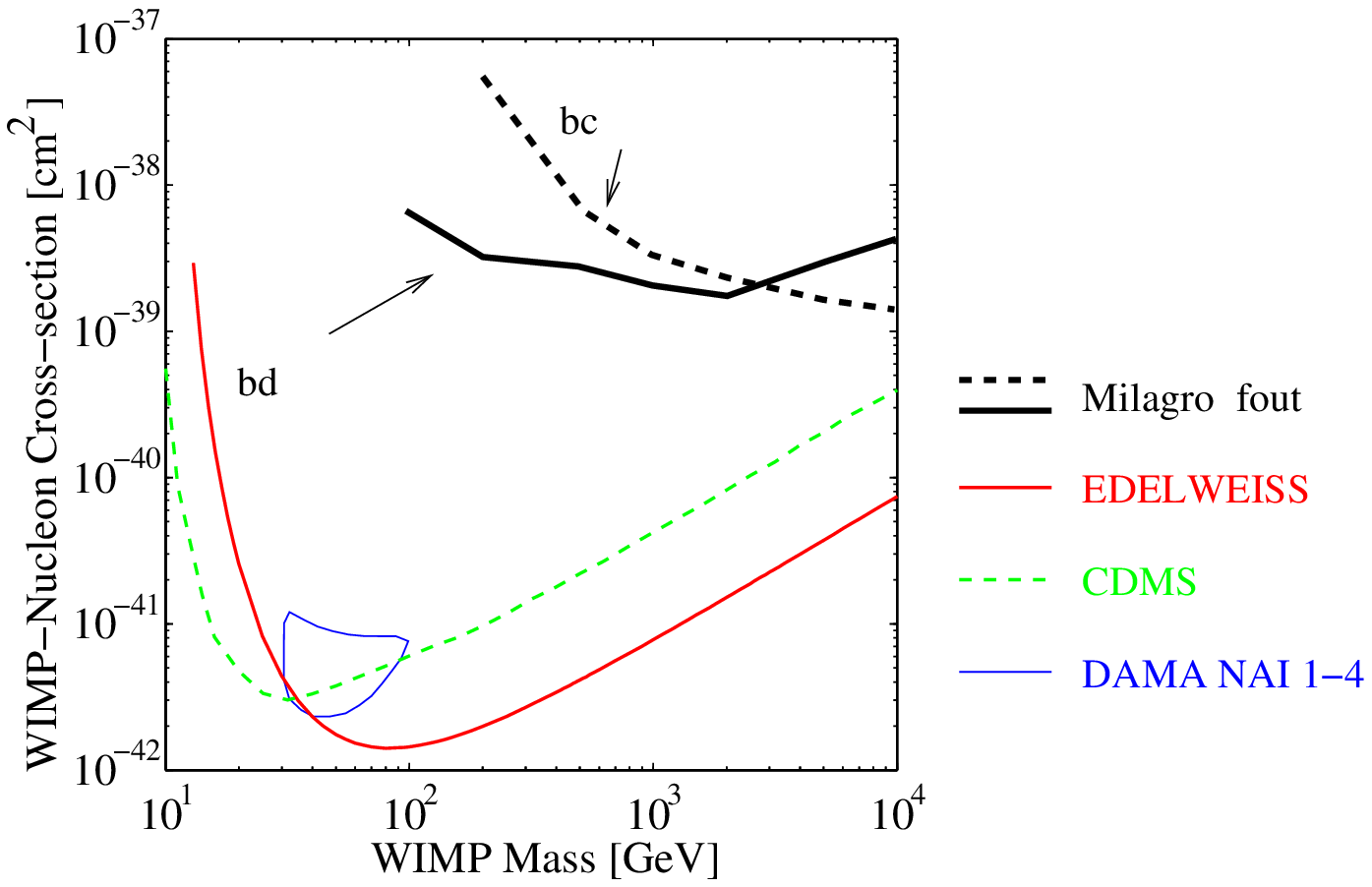}
%\plotone{FIGURES/my_edelweiss.eps}
\caption{Upper boundary of a $90\%$ one-sided confidence interval
         on neutralino-nucleon cross-sections obtained
         from different direct-search experiments. The closed contour is 
         the allowed region at $3\sigma$ confidence level from the DAMA 
         experiment. The plot is adopted from~\cite{dmtools}.
         The two Milagro curves plotted assume an equilibrium 
         situation where the annihilation rate $J(m_{\chi})$ equals the 
         capture rate $I(m_{\chi})$,  and consider the direct and
         nondirect annihilation modes for preset values of 
         $b_{\gamma}$~\cite{ullio,darksusy}. The Milagro limit is 
         obtained by dividing the plotted values by the fraction of 
         annihilations that occur outside the Sun, $f_{out}$,
         for which calculations vary from $10^{-1}$ to 
         $10^{-16}$~\cite{Strausz,Hooper,LazarPhD}.
         \label{fig:edelweiss}}
\end{figure}

The interpretation of the constructed limit on the gamma-ray flux is
highly model dependent. It is based, for instance, on assumptions
regarding the shape of the velocity distribution of the dark matter in
the galactic halo and its density profile in the Solar System. 
Therefore, several assumptions are made to construct limits on
physically interesting quantities.

We assume Maxwellian velocity distribution of the neutralinos in the
Solar system with mean velocity $V_{0}=220\ km/s$ and width
$2V_{0}^{2}$. The photon flux of equation~(\ref{eq:Begstrom_flux}) at
the Earth due to neutralino annihilations can be computed as:

\begin{equation}
   dF_{\chi}(E_{\gamma})= f_{out}(m_{\chi})\cdot f_{e}\cdot
        b_{\gamma}(E_{\gamma},m_{\chi})\cdot 
                 \frac{J(m_{\chi})}{4\pi L_{\oplus}^{2}}\ dE_{\gamma}
\label{eq:flux_general}
\end{equation}

%%\[ d F_{\chi}(E_{\gamma})=
%%            \frac{\rho_{0}}{0.3\ (GeV/cm^{3})}\cdot
%%            \frac{\sigma_{p\chi}}{10^{-43}\ (cm^{2})}\cdot 
%%                    b_{\gamma}(E_{\gamma},m_{\chi})\cdot
%%                      f_{out}(m_{\chi})\cdot
%%  \frac{J(m_{\chi})}{5.6\cdot 10^{27}\ (s^{-1})}\ dE_{\gamma} \ \ \ \ 
%% (cm^{-2}s^{-1}) \]

where $b_{\gamma}(E_{\gamma},m_{\chi})$ is differential photon yield per
neutralino for producing a photon with energy $E_{\gamma}$ in
neutralino-neutralino annihilation, $f_{out}(m_{\chi})$ is the fraction
of neutralinos annihilating outside the Sun, $f_{e}$ is the fraction of
produced photons which escape from the Sun and is of the order of $1/2$,
$J(m_{\chi})$ is the total neutralino annihilation rate in the Solar
system and $L_{\oplus}=1.5\cdot 10^{11}\ m$ is the mean Sun-Earth
distance.

Given the functional form of the flux from
equation~(\ref{eq:Begstrom_flux}), the photon yield
$b_{\gamma}(E_{\gamma},m_{\chi})$ is:

\begin{equation}
   b_{\gamma}(E_{\gamma},m_{\chi}) =
    b_{\gamma}^{\delta}(m_{\chi})\delta\Big(E_{\gamma}-m_{\chi}\Big) +
     \frac{b_{\gamma}^{c}(m_{\chi})}{m_{\chi}}\cdot
     \frac{\Big(\frac{E_{\gamma}}{m_{\chi}}\Big)^{-3/2}e^{-7.8E/m_{\chi}}}%
                                {\int_{0.01}^{1}x^{-3/2}e^{-7.8x}dx}
\label{eq:yield}
\end{equation}

where $b_{\gamma}^{\delta}$ ($b_{\gamma}^{c}$) is the number of photons
produced per annihilation directly (indirectly).

One can also make an assumption that an equilibrium situation has been
reached and that the annihilation rate $J(m_{\chi})$ and the capture
rate $I(m_{\chi})$ are identical. (We use this in
figure~\ref{fig:edelweiss} below.)

A 3-D calculation has been performed~\cite{LazarPhD} to determine the
rate $I(m_{\chi})$ of WIMP capture by the Sun as a function of the
neutralino mass.  For given local galactic dark matter density
$\rho_{0}$, a structure-less\footnote{Structure-less scattering is the
one with no restrictions other than energy and momentum conservation.} 
$\chi-p$ elastic cross-section $\sigma_{p\chi}$ determines how often a
neutralino passing through the Sun scatters and loses enough energy to
get gravitationally captured. The results of this calculation are
presented in table~\ref{table:predicted_capture}. Since $I(m_{\chi})$ is
proportional to $\rho_{0}\sigma_{p\chi}$, it will be normalized to the
capture rate computed at $\rho_{0}=0.3\ GeV/cm^{3}$ and
$\sigma_{p\chi}=10^{-43}\ cm^{2}$.

%%%%%$I(m_{\chi})=\frac{\rho_{0}}{0.3GeV/cm^{3}}
%%%%%\frac{\sigma_{p\chi}}{10^{-43}cm^{2}}I_{0}(m_{\chi})$. 

A limit on $\sigma_{p\chi}b_{\gamma}$ would provide constraints on
parameters of Super Symmetric models. Using the formulae
(\ref{eq:flux_constraints},\ref{eq:flux_general},\ref{eq:yield}),
however, one obtains neutralino-mass-dependent limits on the product of
$\rho_{0}\sigma_{p\chi}b_{\gamma}f_{out}(J/I)$ as presented in table
\ref{table:neutralino_limits} and figure~\ref{fig:neutralino_limits}.
While the value of the local dark matter density $\rho_{0}$ is known
relatively well, there are substantial disagreements on the fraction of
neutralino annihilations near the Sun $f_{out}$.

The problem of WIMP capture on bound near-solar orbits was considered in
\cite{PressSpergel,Gould1987}. To our knowledge, the first
one-dimensional computer simulation of the distribution of the
annihilation points near the Sun was treated in \cite{Strausz}. There,
it was assumed that the Solar system consists of a uniform density Sun
only and that the capture of a particle happens in its first scattering
inside the Sun with an additional assumption that the Solar system has
reached a dynamic equilibrium and that the capture rate $I(m_{\chi})$ is
equal to the annihilation one $J(m_{\chi})$. \cite{Strausz} concluded
that the fraction of all annihilations happening outside the Sun is
$f_{out}\sim 10^{-5}-10^{-7}$. Under essentially the same assumptions, a
simulation done by \cite{Hooper} provided a drastically different
prediction of $f_{out}\sim 10^{-14}-10^{-16}$. However, a 3-D computer
simulation of the neutralino annihilation distribution~\cite{LazarPhD}
provides an estimate $f_{out}\sim 10^{-1}$.

Therefore, in figure~\ref{fig:edelweiss} we illustrate $90\%$ one-sided
confidence intervals for two different simplified cases: first
$b_{\gamma}^{\delta} = 0.001$ and $b_{\gamma}^{c} = 0$, and second
$b_{\gamma}^{\delta} = 0$ and $b_{\gamma}^{c} = 1$.  In both cases we
take $\rho_{0}=0.3\ GeV/cm^{3}$, $J(m_{\chi}) = I(m_{\chi})$, and
$f_{out} = 1$.  It should be noted that while $b_{\gamma}^{\delta}$ is
always less than unity~\cite{ullio}, $b_{\gamma}^{c}$ can be as high as
$10$ for some Super Symmetric models~\cite{darksusy}.

% ===================================================================

\section{Conclusion}

This work presents the first attempt to detect TeV photons produced by
annihilating neutralinos captured into the Solar system. Analysis of the
Milagro data set collected during 2000-2001 shows no evidence for a
gamma-ray signal due to such a process. The limit on the possible
gamma-ray flux due to such a process with significance $2.9\cdot
10^{-7}$ and the power $97.7\%$ has been set (see
table~\ref{table:flux_constraints} and
figure~\ref{fig:flux_constraints}). Even in the absence of a clear
signal the constructed exclusion limit may constrain the values of free
parameters of supersymmetric models (see
table~\ref{table:neutralino_limits} and
figure~\ref{fig:neutralino_limits}). In addition, a standard $90\%$
one-sided confidence interval on the magnitude of the photon flux due to
near-Solar neutralino annihilations has been constructed.

The interpretation of the constructed limit on the gamma-ray flux is
highly model dependent. Conversion of the flux measurement to a cross
section limit requires knowledge of the annihilation rate,
$J(m_{\chi})$, the fraction of annihilations outside the Sun, $f_{out}$,
and photon yield per annihilation, $b_{\gamma}$. These, in turn, depend
on parameters of astrophysical and Super Symmetric models. As mentioned
in the paper, some of these can be estimated using simple assumptions.
For example, one may assume an equilibrium situation when capture rate
is equal to the annihilation one. However, absence of a reliable
estimate on the fraction of annihilations happening outside the Sun
makes it hard to interpret the limits in terms of the theoretically
interesting $b_{\gamma}\sigma_{p\chi}$. Therefore, the results are
presented with all these parameters written out explicitly.

Continuous improvements in reconstruction algorithms, detector
modifications and longer observation times will lead to a better upper
limit. One of the factors which led to a deterioration of the
constructed upper limit is the inability to compensate for presence of
the Solar cosmic-ray shadow due to the intricate structure of the Solar
magnetic fields. Once the cosmic-ray shadow of the Sun is understood
quantitatively, it may be possible to improve upon the limit.

%a better upper limit can be constructed.

%The performed observation allows setting a limit on other possible 
%processes which may produce high energy photons near the Sun such as 
%albedo?????

% ===================================================================

% ===================================================================

\begin{acknowledgments}

We acknowledge Scott Delay and Michael Schneider for their dedicated
efforts in the construction and maintenance of the Milagro experiment. 
This work has been supported by the National Science Foundation (under
grants
PHY-0070933, %UCI
-0075326, %Milagro Operations
-0096256, %UW-Madison
-0097315, %LANL via UMD 
-0206656, %NYU current; previous is PHY-9901496
-0245143, %UCSC; previous is PHY-0070927
-0302000, %UMD
and
ATM-0002744) %UNH
the US Department of Energy (Office of High-Energy Physics and Office of
Nuclear Physics), Los Alamos National Laboratory, the University of
California, and the Institute of Geophysics and Planetary Physics.

\end{acknowledgments}

% ===================================================================

\bibliography{milagro_wimps}
\bibliographystyle{plain}

\end{document}